\documentclass[conference]{IEEEtran}
\IEEEoverridecommandlockouts

\usepackage[
    paperwidth=8.5in,
    paperheight=11in,
    left=0.673in,
    right=0.673in,
    top=1.047in,
    bottom=1.0in,
    columnsep=0.166in
]{geometry}

\usepackage{cite}
\usepackage{amsmath,amssymb,amsfonts}
\usepackage{mathtools}
\usepackage{algorithmic}
\usepackage{graphicx}
\usepackage{textcomp}
\usepackage{xcolor}

\usepackage{algorithm}
\usepackage{amsthm}
\theoremstyle{definition}

\newtheorem{remark}{Remark}
\theoremstyle{plain}

\newtheorem{theorem}{Theorem}
\newtheorem{lemma}{Lemma}
\newtheorem{corollary}{Corollary}

\theoremstyle{remark}
\newtheorem*{proof_sketch}{Proof sketch}

\newcommand{\R}{\mathbb{R}}
\newcommand{\E}{\mathbb{E}}

\usepackage{booktabs}
\usepackage{array}
\usepackage{multirow}
\usepackage{multicol}

\usepackage{makecell}

\usepackage{cuted}

\def\BibTeX{{\rm B\kern-.05em{\sc i\kern-.025em b}\kern-.08em
    T\kern-.1667em\lower.7ex\hbox{E}\kern-.125emX}}
\begin{document}

\title{A PAC-Bayesian Analysis of Channel-Induced Degradation in Edge Inference
}

\author{\IEEEauthorblockN{Yangshuo He and Guanding Yu}
\IEEEauthorblockA{College of Information Science \& Electronic Engineering\\
Zhejiang University,
Hangzhou, China \\
Email: \{sugarhe, yuguanding\}@zju.edu.cn}
\and
\IEEEauthorblockN{Jingge Zhu}
\IEEEauthorblockA{Department of Electrical and Electronic Engineering \\
The University of Melbourne,
Melbourne, Australia \\
Email: jingge.zhu@unimelb.edu.au}
}

\maketitle

\begin{abstract}
In the emerging paradigm of edge learning, neural networks (NNs) are partitioned across distributed edge devices that collaboratively perform inference via wireless transmission. 
However, deploying NNs for edge inference over wireless channels inevitably leads to performance degradation, as the exact channel realizations in the inference stage are not known in the training stage.
In this paper, we establish a theoretical framework to evaluate and bound this performance degradation. Inspired by statistical learning theory, we define a wireless generalization error to characterize the gap between the empirical performance during training and the expected inference performance under the true stochastic channel. 
To enable theoretical analysis, we introduce an augmented NN model that incorporates channel statistics directly into the weight space. Leveraging the PAC-Bayesian framework, we derive a high-probability bound on this error, which provides theoretical guarantees for wireless inference performance.
Furthermore, we propose a channel-aware training algorithm that minimizes a tractable surrogate objective based on the derived bound. Simulations demonstrate that the proposed algorithm effectively improves wireless inference performance and model robustness under various channel conditions.
\end{abstract}

\section{Introduction}
The emerging demand for deploying Artificial Intelligence applications on edge devices has motivated the integration of deep learning and wireless communications, termed edge AI \cite{edge-AI}.
A key application in this area is edge inference, specifically via an edge co-inference architecture \cite{wireless-distributed-learning-survey}. In this setting, a neural network (NN) is split and deployed across separate edge devices, where the transmitter processes initial NN layers and sends extracted features over wireless channels to the receiver \cite{edge-inference}. 
The deployed NNs in edge inference are generally trained using specific datasets and learning algorithms.
However, a critical challenge arises during actual edge deployment: the transmitted features are corrupted by stochastic wireless channels. Even if a model is trained using the exact statistics of the deployment environment, the instantaneous channel realizations encountered during edge inference remain unknown and will inevitably differ from those seen during training. This fundamental mismatch creates an unavoidable performance gap between the empirical accuracy observed during training and the true inference performance.

To address this challenge, recent advancements in edge learning often utilize end-to-end joint source-channel coding. These works jointly optimize NN-based encoder and decoder under a presumed static channel condition \cite{JSCC,JSCC-DeepJSCC}. 
Recent efforts explore adaptive NN architectures to handle varying inference channels, such as variable feature-length schemes \cite{edge-IB} and adaptive multi-level feature transmission \cite{JSCC-multilevel}. 
Alternative approaches include vector quantization encoders \cite{JSCC-VQ} and self-attention mechanisms that leverage the intrinsic robustness of NNs \cite{blind-JSCC}.
Other works explicitly utilize channel state information (CSI) through attention mechanisms to adjust source encoder's compression rate \cite{JSCC-attention}, and dynamic feature selection using real-time CSI \cite{JSCC-multitask}. While effective, these methods either remain vulnerable to inaccurate real-time CSI estimation or lack a rigorous theoretical foundation guaranteeing inference performance under inherent channel stochasticity.

While theoretical analysis of inference degradation over stochastic channels remains limited, we are inspired by research in statistical learning theory. Specifically, these studies model weight quantization as a stochastic transition and bound the performance degradation due to weight compression \cite{compression,variable-compression}. Motivated by these works, we formalize the edge inference performance deterioration as a wireless generalization error.
Unlike the traditional generalization regarding unseen data samples, we extend this concept to characterize the model's robustness against unseen channel conditions.

\begin{figure*}[htbp]
  \centering
  \includegraphics[width=0.9\textwidth]{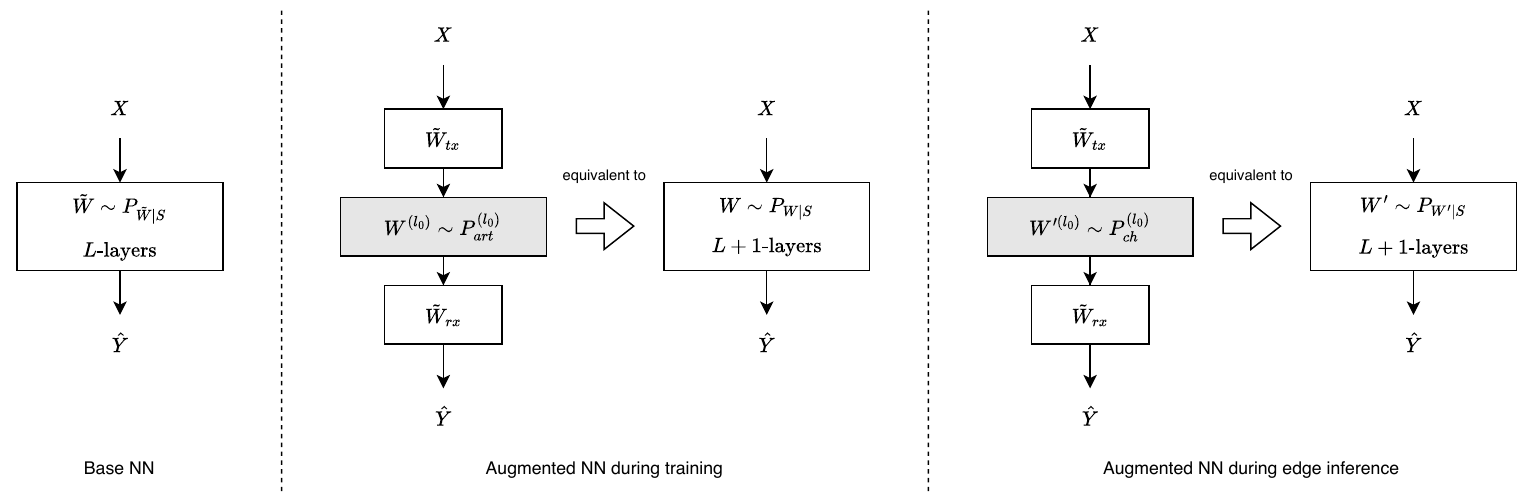}
  \caption{\textbf{Left}: the $L$-layer base NN, weights $\tilde{W}$ are given by the learning algorithm $P_{\tilde{W}|S}$. \textbf{Middle}: the augmented $L+1$-layer NN in training phase, where the additional $l_0$-th layer represents an artificial channel. Weights $W$ are given by the learning algorithm $P_{W|S}$ defined in \eqref{equ:learning_algorithm}. \textbf{Right}: the augmented $L+1$-layer NN deployed for edge inference, where the additional $l_0$-th layer represents a channel in actual environment. Weights $W'$ are given by the conditional distribution $P_{W'|S}$ defined in \eqref{equ:P_W'|S}.}
  \label{fig:system_model}
  \vspace{-0.5cm}
\end{figure*}

To facilitate theoretical analysis, we introduce an augmented NN model that explicitly integrates channel statistics into weight space. Specifically, we model the wireless channel as an additional $l_0$-th layer in NN, with non-learnable stochastic weights. During inference, the weight distribution of this layer is determined by the physical channel statistics. In contrast, during training, this distribution is a component of the learning algorithm which may be controlled or simulated by the algorithm designer. Other layers' weights are optimized during training phase and frozen during the inference phase. We model the performance degradation due to unseen channel realizations during inference as a wireless generalization error. Leveraging the PAC-Bayesian framework \cite{pac-bayes-catoni,pac-bayes-mcallester}, we derive high-probability bounds that explicitly quantify this error.
Finally, we propose a channel-aware training algorithm designed to minimize a tractable objective based on the derived bound. Simulations demonstrate that the derived bounds provide theoretical guarantees for edge inference and our algorithms mitigate performance degradation caused by wireless channels.

\section{Problem Formulation}
\label{sec:problem_formulation}

The primary focus of this work is to derive a theoretical upper bound on the generalization error of NNs deployed in wireless edge systems. 
We first introduce an augmented NN model that characterizes wireless channels as an NN layer with stochastic weights. Leveraging a PAC-Bayesian framework, we then quantify the performance degradation induced by this inherent channel randomness.

\subsection{System Model and Augmented Network}
\label{sec:augmented_network}
Let $\tilde{W}=(\tilde{W}^{(1)}, \tilde{W}^{(2)},\cdots,\tilde{W}^{(L)})\in\tilde{\mathcal{W}}$ be the weights of a base $L$-layer NN. We consider an instance space $\mathcal{Z}=\mathcal{X}\times\mathcal{Y}$, which is the product of a feature space $\mathcal{X}$ and a label space $\mathcal{Y}$. A training set $S=(Z_1,Z_2,\cdots,Z_n)$ consists of $n$ i.i.d. samples $Z_i\in\mathcal{Z}$ drawn from a data distribution $P_Z$. We denote the distribution of the training set $S$ as $P_S=P_Z^n$.

When an NN task is deployed in wireless systems, the NN is generally split into two components that communicate through the wireless channel: a transmitter part $\tilde{W}_{tx}$ deployed on the local device and a receiver part $\tilde{W}_{rx}$ deployed on the edge server. Due to channel impairments, the hidden layer features transmitted by the NN on the transmitter may be distorted, causing errors during the subsequent inference in NN on the receiver. 
These channel-induced distortions on the features can be captured by a fully connected (FC) layer. Therefore, we introduce an additional layer to characterize the NN deployed in wireless environments, as shown in Figure \ref{fig:system_model}. 

This structural equivalence allows us to integrate the channel statistics directly into the weight space.
Let $l_0$ denote the additional layer, and the layer output can be formulated as $f^{(l_0)} = M^{(l_0)}f^{(l_0-1)} + B^{(l_0)}$, where $f^{(l_0-1)},f^{(l_0)}\in\R^d$ denote the input and output features, respectively. The weights of this layer $W^{(l_0)}=(M^{(l_0)},B^{(l_0)})$ consist of a weight matrix $M^{(l_0)}\in\R^{d\times d}$ and a bias vector $B^{(l_0)}\in\R^d$. These stochastic weights are determined by the specific channel realizations.
While this formulation represents an affine transformation, it acts as a versatile abstraction capable of modeling various wireless environments. In this framework, simple modeling of non-linear operations in digital channels like quantization and coding can be encapsulated into the effective stochastic weights. 

By incorporating this additional channel layer, we define an augmented NN consisting of $L+1$-layer weights. This structural equivalence combines the learnable base weights $\tilde{W}$ and the non-learnable stochastic channel weights $W^{(l_0)}$ into a weight space $W=(\tilde{W},W^{(l_0)})\in\mathcal{W}$. Specifically, $(W^{(l)})_{l<l_0}$ correspond to the transmitter part $\tilde{W}_{tx}$ while $(W^{(l)})_{l>l_0}$ correspond to the receiver part $\tilde{W}_{rx}$. In the rest of this article, we will focus on the $L+1$-layer augmented NN.

\subsection{Probabilistic framework and Wireless Generalization Error}
\label{sec:probabilistic}
Unlike traditional settings that assume identical weight realizations during training and inference, we explicitly define distinct distributions for the augmented NN's weights across two phases.
Let $W$ and $W'$ denote the weights of the augmented NN during the training phase and the inference phase, respectively. 
During training phase, we define the $l_0$-th layer weights as an algorithmic design choice rather than a physical constraint. These weights are drawn from a fixed ``artificial channel'' distribution $W^{(l_0)}\sim P_{art}^{(l_0)}$. This distribution is chosen as part of the learning algorithm and can be configured to represent a perfect, noiseless channel or calibrated to match the true inference channel statistics.
The overall learning algorithm is formalized as a stochastic mapping from the training set $S$ to the augmented weights $W$, characterized by conditional distribution $P_{W|S}$ over $\mathcal{W}$:
\begin{equation}
    \label{equ:learning_algorithm}
    \begin{split}
        P_{W|S}(w|s) & \coloneqq P_{\tilde{W}|W^{(l_0)}S}\left(\tilde{w}|w^{(l_0)},s\right)P_{art}^{(l_0)}\left(w^{(l_0)}\right),
    \end{split}
\end{equation}
where $P_{\tilde{W}|W^{(l_0)}S}$ characterizes dependence of the learnable base weights on both the designed stochastic layer and dataset.

When the NN is deployed in a wireless system for edge inference, the base weights $\tilde{W}$ are frozen and inherited directly from the trained weights $W$, implying $\tilde{W}'=\tilde{W}$. However, the physical channel during inference is stochastic and determined by the actual wireless environment. We formulate this inference channel as weights following a distribution ${W'}^{(l_0)}\sim P_{ch}^{(l_0)}$. Thus, for given trained weights $w$, the edge inference weights follow a conditional distribution as follows
\begin{equation}
    \label{equ:P_W'|W}
    \begin{split}
        P_{W'|W}(w'|w) \coloneqq \delta_{\tilde{w}}\left(\tilde{w}'\right)P_{ch}^{(l_0)}\left({w'}^{(l_0)}\right),
    \end{split}
\end{equation}
where $\delta_{\tilde{w}}$ is the Dirac measure centered at the trained base weights.
Given trained weights $W$, the inference weights $W'$ are conditionally independent of the training set $S$. 
Therefore, the joint distribution of $W',W$ given $S$ is $P_{W'W|S}=P_{W'|W}P_{W|S}$ and the marginal distribution over the inference weights induced by $P_{W'W|S}$ is derived as
\begin{equation}
    \label{equ:P_W'|S}
    \begin{split}
        P_{W'|S}\left(w'|s\right) = P_{\tilde{W}|S}\left(\tilde{w}'|s\right)P_{ch}^{(l_0)}\left({w'}^{(l_0)}\right),
    \end{split}
\end{equation}
where $P_{\tilde{W}|S}$ is the marginal distribution induced by $P_{W|S}$.
Crucially, even if the learning algorithm is designed such that the artificial training channel distribution matches the inference channel distribution ($P_{art}^{(l_0)}=P_{ch}^{(l_0)}$), the instantaneous realizations $w$ and $w'$ will be different.

To evaluate the robustness of the NN against unseen channels, we quantify the performance degradation using a loss function $l:\mathcal{W}\times \mathcal{Z}\to \R^+$. 
For specific weights $w$, the inference quality is measured by the population risk on the sample space $L(w) \coloneqq \E_{P_Z}[l(w,Z)]$. However, since the data distribution $P_Z$ is generally unknown, the population risk $L(w)$ is unobservable. The observable training performance is given by the empirical risk $\hat{L}(w,S) \coloneqq \frac{1}{n}\sum_{i=1}^nl(w,Z_i)$.

In the PAC-Bayesian framework, the learning algorithm is stochastic (e.g., the weights are obtained by stochastic gradient descent using the training data). Therefore, the true inference quality is measured by the expected population risk $\mathbb{E}_{P_{W'|S}}[L(W')]$. For a given training set $S$, the expected empirical risk is $\E_{P_{W|S}}[\hat{L}(W,S)]$. Based on these definitions, we define the performance degradation induced by wireless edge deployment as wireless generalization error:
\begin{equation}
    \label{equ:wireless_generalization_gap}
    \Delta \coloneqq \mathbb{E}_{P_{W'|S}}\left[L\left(W'\right)\right] - \mathbb{E}_{P_{W|S}}\left[\hat{L}\left(W,S\right)\right].
\end{equation}
This difference mathematically captures the discrepancy between the observed empirical performance during training and the true performance in edge inference. 
Moreover, since $\mathbb{E}_{P_{W'|S}}[L(W')] = \mathbb{E}_{P_{W|S}}[\hat{L}(W,S)]+\Delta$, providing an upper bound for $\Delta$ inherently yields a performance guarantee for edge inference performance.
Next, we will derive PAC-Bayesian bounds for this difference, which characterizes the inherent robustness against wireless channel perturbations.

\section{Main Results}
\label{sec:main_results}
In this section, we derive PAC-Bayesian bounds for the wireless generalization error defined in \eqref{equ:wireless_generalization_gap}.
It connects the expected population risk under the stochastic wireless channel during inference to the expected empirical risk observed during the training phase.
We begin by deriving a general bound for the scenario where the NN is trained with the channel layer $W^{(l_0)}$ generated by an artificial channel. Noting that the bound becomes vacuous in the special case when the artificial channel is perfect, we provide an alternative bound based on a Lipschitz continuity assumption for this special case. Proofs are postponed to Appendix \ref{apx:proof_theorem_1}-\ref{apx:proof_corollary_1}.

\begin{theorem}
    \label{thm:simple_channel_bound}
    Assume that the loss function $l(w,z)$ is $\sigma$-sub-Gaussian under $P_Z$ for all $w\in\mathcal{W}$ and is $\sigma_0$-sub-Gaussian under $P_{W'|S}$ for all $z\in\mathcal{Z}$, then for all data-independent prior distribution $Q_{\tilde{W}}$ over $\tilde{\mathcal{W}}$, w.p. at least $1-\epsilon$ over $S$, we have
    \begin{equation}
        \label{equ:simple_channel_bound}
        \begin{split}
            & \Delta \leq \sqrt{2\sigma_0^2\left(I_S\left(\tilde{W};W^{(l_0)}\right) + D\left(P_{art}^{(l_0)}\big\|P_{ch}^{(l_0)}\right)\right)} \\
            &  + \sqrt{\frac{2\sigma^2\left(\E_{P_{art}^{(l_0)}}\left[D\left(P_{\tilde{W}|W^{(l_0)}S}\|Q_{\tilde{W}}\right)\right]+\log\frac{\sqrt{n}}{\epsilon}\right)}{n-1}},
        \end{split}
    \end{equation}
    where $I_S(\tilde{W};W^{(l_0)}) \coloneqq \E_{P_{art}^{(l_0)}}[D(P_{\tilde{W}|W^{(l_0)}S}\|P_{\tilde{W}|S})]$. 
\end{theorem}

\begin{proof_sketch}
    The proof proceeds in three steps. First, adding and subtracting $\E_{P_{W|S}}[L(W)]$ decomposes the wireless generalization error $\Delta$ into the expected risk difference between $P_{W'|S}$ and $P_{W|S}$, and the standard generalization gap of $P_{W|S}$. Second, applying the Donsker-Varadhan lemma alongside the sub-Gaussian assumptions upper bounds both components, yielding terms governed by the KL divergences $D(P_{W|S}\|P_{W'|S})$ and $D(P_{W|S}\|Q_W)$, respectively. Finally, we obtain the stated bound by simplifying these divergences using the probabilistic definitions from Section \ref{sec:probabilistic}.
\end{proof_sketch}

\begin{remark}
    Theorem \ref{thm:simple_channel_bound} provides a fundamental decoupling of wireless generalization error into two components: the first term captures the penalty induced by the stochastic nature of wireless channels, while the second term represents the standard generalization bound scaling with $\mathcal{O}(1/\sqrt{n})$ \cite{genearlization-bound-book}. 
    Specifically, KL divergence $D(P_{art}^{(l_0)}\|P_{ch}^{(l_0)})$ explicitly captures the cost of statistical discrepancy between the artificial channel distribution in training and the true physical channel distribution encountered during inference.
    Meanwhile, the mutual information $I_S(\tilde{W};W^{(l_0)})$ measures the information about specific channel realizations generated during training contained in NN's weights. Since the forward pass in training depends on the specific drawn weights $w^{(l_0)}$, the resulting gradient updates couple the final base weights $\tilde{W}$ to the specific realization. This reveals a critical trade-off in wireless edge learning. 
    An NN that perfectly adapts to these channel realizations will exhibit high mutual information. While this can minimize the empirical risk, it also indicates overfitting, leaving the model to suffer performance degradation caused by unknown and instantaneous realizations during actual edge deployment. 
\end{remark}

\begin{remark}
    Remarkably, even when the artificial channel statistics during training align with the true inference channel (i.e., $P_{art}^{(l_0)}=P_{ch}^{(l_0)}$) and $D(P_{art}^{(l_0)}\|P_{ch}^{(l_0)})=0$, our bound does not simply reduce to the conventional PAC-Bayesian bound. This highlights a fundamental paradigm shift arising from our formulation of edge inference. 
    Conventional PAC-Bayes assumes the identical training and inference weight realizations, 
    i.e., $P_{W'|W}(w'|w)=\delta_{w}(w')$. However, in the wireless edge learning scenario, only the base model weights $\tilde{W}$ are frozen and identical, while the weights of the channel layer are dynamically re-sampled from the underlying distributions during both training and inference. 
    Treating the wireless channel as a stochastic, non-learnable layer whose exact realization remains unknown during edge inference creates a crucial difference from standard machine learning architectures.
\end{remark}

Consider a simple scenario where the model is trained in an ideal, channel-free condition, without considering future edge deployment (i.e., $P_{art}^{(l_0)}$ is a Dirac measure centered at the identity matrix and zero vector). Consequently, the mutual information $I_S(\tilde{W};W^{(l_0)})$ becomes zero and $\E_{P_{art}^{(l_0)}}[D(P_{\tilde{W}|W^{(l_0)}S}\|Q_{\tilde{W}})]$ simplifies to $D(P_{\tilde{W}|S}\|Q_{\tilde{W}})$. 
However, if the channel distribution $P_{ch}^{(l_0)}$ is defined over a continuous random variable (e.g., Rayleigh fading), the KL divergence $D(P_{art}^{(l_0)}\|P_{ch}^{(l_0)})$ becomes infinite, rendering the bound vacuous. 
To address this extreme yet practically common case of deploying traditionally trained models over stochastic edge systems, we introduce a Lipschitz continuity assumption, leading directly into Theorem \ref{thm:perfect_training_simple_channel_bound}.

\begin{theorem}
    \label{thm:perfect_training_simple_channel_bound}
    Suppose the NN is trained under a perfect channel, i.e., $P_{art}^{(l_0)}(w^{(l_0)})=\delta_{(\mathbf{I},\mathbf{0})}(w^{(l_0)})$, and deployed for edge inference under a channel $P_{ch}^{(l_0)}(w^{(l_0)})$. Assume that the loss function $l(w,z)$ is $\sigma$-sub-Gaussian under $P_Z$ for all $w\in\mathcal{W}$, and is $K$-Lipschitz continuous on $\mathcal{W}$ for all $z\in\mathcal{Z}$, then, for all data-independent prior $Q_{\tilde{W}}$ over $\tilde{\mathcal{W}}$, with probability at least $1-\epsilon$ over $S$, we have
    \begin{equation}
        \label{equ:perfect_training_simple_channel_bound}
        \begin{split}
            \Delta & \leq \E_{P_{ch}^{(l_0)}}\left[K\left\|{W'}^{(l_0)}-(\mathbf{I},\mathbf{0})\right\|_F\right] \\
            & \quad + \sqrt{\frac{2\sigma^2}{n-1}\left(D(P_{\tilde{W}|S}\|Q_{\tilde{W}})+\log\frac{\sqrt{n}}{\epsilon}\right)}.
        \end{split}
     \end{equation}
\end{theorem}

While Theorem \ref{thm:perfect_training_simple_channel_bound} provides a theoretical guarantee for the wireless generalization error, computing the exact expectation over $P_{ch}^{(l_0)}$ is often intractable for complex fading channels with spatial correlation. Although numerical methods (e.g., Monte Carlo) can approximate this term, they lack rigorous theoretical guarantees. 
Noting that the channel-induced penalty in this formulation remains a constant during training and thus ineffective for algorithm designs, our primary goal here is to establish a theoretically tractable, closed-form guarantee for wireless edge inference. 
To achieve this when evaluating over the exact physical channel is intractable, we can introduce a simplified, data-independent prior $Q^{(l_0)}$ over $\mathcal{W}^{(l_0)}$ to derive an analytic relaxation, as shown in the following corollary.

\begin{corollary}
    \label{thm:perfect_training_simple_channel_bound_relaxed}
    Suppose the NN is trained under a perfect channel. Assume that the loss function $l(w,z)$ is $\sigma$-sub-Gaussian under $P_Z$ for all $w\in\mathcal{W}$ and is $K$-Lipschitz continuous on $\mathcal{W}$ for all $z\in\mathcal{Z}$, then, for all data-independent prior distribution $Q_{\tilde{W}}$ and $Q^{(l_0)}$, w.p. at least $1-\epsilon$ over $S$, we have
    \begin{equation}
        \label{equ:perfect_training_simple_channel_bound_relaxed}
        \begin{split}
            \Delta & \leq \log \E_{Q^{(l_0)}}\left[e^{K\left\|{W'}^{(l_0)}-(\mathbf{I},\mathbf{0})\right\|_F}\right] + D\left(P_{ch}^{(l_0)}\|Q^{(l_0)}\right) \\
            & \quad + \sqrt{\frac{2\sigma^2}{n-1}\left(D(P_{\tilde{W}|S}\|Q_{\tilde{W}})+\log\frac{\sqrt{n}}{\epsilon}\right)}.
        \end{split}
    \end{equation}
\end{corollary}

\section{Channel-aware Training Algorithm}
\label{sec:algorithm}
Inspired by the theoretical analysis, we propose a channel-aware training algorithm that explicitly utilizes channel statistics to enhance edge inference performance. 
Instead of considering the wireless channel as simply a distortion on hidden layer features and applying empirical risk minimization (ERM), our method explicitly incorporates channel stochasticity into the learning objective. 
Specifically, given the definition in \eqref{equ:wireless_generalization_gap}, we can formulate a tractable upper bound on the expected population risk during edge inference by combining the expected empirical risk with our derived bound for $\Delta$. 
Therefore, penalizing the upper bound for wireless generalization error acts as a principled regularization mechanism.
It prevents the model from overfitting to specific channel realizations observed in training, thereby enhancing robustness against unknown channel conditions during edge inference. 

To achieve this, we formulate a surrogate objective function combining both the expected empirical risk and the derived upper bound.
While standard algorithms optimize all NN weights, the modeling in edge learning designates the $l_0$-th layer as a stochastic non-learnable artificial layer during training. Consequently, the optimization is restricted to the learnable base weights $\tilde{W}$. The primary goal of the algorithm is to find a posterior $P_{\tilde{W}|W^{(l_0)}S}$ that minimizes the regularized objective function. By introducing a scaling factor $\eta>0$ for effective optimization, as in \cite{BBB}, we formulate the surrogate objective function as follows
\begin{equation}
    \label{equ:objective}
    \begin{split}
        & \mathcal{J}\big(P_{\tilde{W}|W^{(l_0)}S}\big) = \mathbb{E}_{P_{W|S}}\big[\hat{L}(W,S)\big]  \\
        &+ \eta\sqrt{2\sigma_0^2\Big(I_S\big(\tilde{W};W^{(l_0)}\big) + D\big(P_{art}^{(l_0)}\big\|P_{ch}^{(l_0)}\big)\Big)} \\
        &+ \eta\sqrt{\frac{2\sigma^2\Big(\mathbb{E}_{P_{art}^{(l_0)}}\big[D\big(P_{\tilde{W}|W^{(l_0)}S}\|Q_{\tilde{W}}\big)\big]+\log\frac{\sqrt{n}}{\epsilon}\Big)}{n-1}}.
    \end{split}
\end{equation}
Note that $P_{art}^{(l_0)}$ acts as a fixed, configurable component during training. It can be configured as a perfect, noiseless channel as in the traditional machine learning without considering future edge deployment. Alternatively, within an end-to-end learning framework where the artificial channel layer can be simulated and controlled, knowing the true physical channel statistics $P_{ch}^{(l_0)}$ allows us to minimize the KL divergence $D(P_{art}^{(l_0)}\|P_{ch}^{(l_0)})$. Importantly, even under the ideal strategy that perfectly aligns the artificial channel statistics with true wireless channel statistics (i.e., $P_{art}^{(l_0)}=P_{ch}^{(l_0)}$), the inherent randomness of the instantaneous realizations still implies an inevitable gap, captured by our derived bound. 

Directly minimizing the exact theoretical bound in Theorem \ref{thm:simple_channel_bound} is computationally challenging due to the mutual information and KL divergence. To enable tractable optimization, we employ the variational approximation in \cite{PBB} where the posterior $P_{\tilde{W}|W^{(l_0)}S}$ is modeled as a diagonal Gaussian distribution over $\tilde{\mathcal{W}}\in\R^m$. Specifically, weights are sampled as $\tilde{W}=\mu+\sigma\odot\epsilon$, where $\epsilon\sim\mathcal{N}\left(\mathbf{0},\mathbf{I}_p\right)$ and $\mu=(\mu_1,\cdots,\mu_m),\sigma=(\sigma_1,\cdots,\sigma_m)\in\R^m$ are learnable mean and deviation, respectively. We parametrize the deviation element-wise as $\sigma=\log(1+e^\rho)$ to ensure non-negativity. For the data-independent prior, we select $Q_{\tilde{W}}$ as a centered isotropic Gaussian $\mathcal{N}(\mathbf{0},\sigma_p\mathbf{I}_m)$, where $\sigma_p\in \R$ represents the initial weight variance. Therefore, we have a closed-form expression for the KL divergence:
\begin{equation}
    D\left(P_{\tilde{W}|W^{(l_0)}S}\|Q_{\tilde{W}}\right) = \sum_{j=1}^m\left(\log \frac{\sigma_p}{\sigma_j}+\frac{\mu_j^2+\sigma_j^2-\sigma_p^2}{2\sigma_p^2}\right).
\end{equation}
We approximate the expectation of KL divergence in the second term of \eqref{equ:simple_channel_bound} via Monte Carlo sampling. Specifically, we draw multiple realizations from $P_{art}^{(l_0)}$, optimize the posterior parameters $\mu$ and $\rho$ for each realization, and average the resulting KL divergences. 

To calculate the mutual information $I_S(\tilde{W};W^{(l_0)})$, it is crucial to distinguish between the conditional distribution $P_{\tilde{W}|W^{(l_0)}S}$ and the marginal $P_{\tilde{W}|S}$. Practically, $P_{\tilde{W}|W^{(l_0)}S}$ represents the internal state of the learning algorithm. 
During training on a dataset $s$, the algorithm explicitly samples a specific artificial channel realization $w^{(l_0)}$, yielding a specific conditional posterior over the base weights. Since the gradient updates depend on the forward pass through this stochastic artificial layer, the optimized base weights are inherently coupled to the drawn channel weights. Conversely, $P_{\tilde{W}|S}$ represents the marginal algorithmic output from an external perspective. It captures the overall distribution of the trained base weights given the dataset $s$, marginalized over all possible realizations of the artificial channel weights. 
The mathematical relation is given by $P_{\tilde{W}|S}(\tilde{w}|s)=\E_{w^{(l_0)}\sim P_{art}^{(l_0)}}[P_{\tilde{W}|W^{(l_0)}S}(\tilde{w}|w^{(l_0)},s)].$
Therefore, this marginal probability density can be approximated by sampling different realizations of $W^{(l_0)}$ and taking the average over the Gaussian density of $P_{\tilde{W}|W^{(l_0)}S}$. Utilizing another Monte Carlo loop over sampling $W^{(l_0)}$, we can approximate the mutual information. Consequently, the objective $\mathcal{J}(P_{\tilde{W}|W^{(l_0)}S})$ is differentiable w.r.t. $\mu$ and $\rho$, allowing efficient optimization via the stochastic gradient descent algorithm. 

When we configure the artificial channel as a noiseless, perfect channel during training, we utilize the bound derived in Theorem \ref{thm:perfect_training_simple_channel_bound} to formulate the objective function. However, the global Lipschitz constant $K\coloneqq\sup_{w\in\mathcal{W},z\in\mathcal{Z}}\|\nabla_w l(w,z)\|_2$ is computationally intractable for NNs due to the supremum over the entire weight and sample space. Alternatively, we approximate $K$ using the tractable local gradient norm of the empirical risk, denoted as $\hat{K}(\tilde{W})=\|\nabla_{W}\hat{L}_S(P_{W|S})\|_2$. Consequently, the channel penalty term becomes coupled with the learning algorithm, allowing us to similarly formulate the surrogate objective function as
\begin{equation}
    \label{equ:objective_perfect}
    \begin{split}
        & \mathcal{J}\left(P_{\tilde{W}|S}\right)  = \mathbb{E}_{P_{W|S}}\left[\hat{L}\left(W,S\right)\right]  + \eta D\left(P_{\tilde{W}|S}\|Q_{\tilde{W}}\right) \\
        & \quad + \eta\hat{K}\left(\tilde{W}\right) \E_{P_{ch}^{(l_0)}}\left[K\left\|{W'}^{(l_0)}-(\mathbf{I},\mathbf{0})\right\|_F\right].
    \end{split}
\end{equation}

\section{Simulation}
\label{sec:simulation}
To illustrate the validity of the derived upper bounds and the effectiveness of the proposed channel-aware training algorithm, we evaluate our framework using a synthesized ``two-moon'' dataset \cite{two-moons}. This dataset consists of 1,000 samples in $\R^2$ perturbed by Gaussian noise of standard deviation 0.3. 
The base NN comprises two FC layers with ReLU activation, where the hidden dimension is $d=64$. To simulate the edge inference system, this NN is partitioned from the middle: the first layer is deployed at the transmitter while the second at the receiver, corresponding to $l_0=2$ in the augmented NN. 
In this simulation, we assume an analog communication model over a standard fading channel subjected to additive white Gaussian noise (AWGN). The received signal is expressed as $\hat{f}=Hf^{(1)}+N$, where $H=\text{diag}(H_1,\cdots,H_d)\in\R^{d\times d}$ is a diagonal matrix with i.i.d. fading gains $H_i\sim \mathcal{N}(\mu_{m},\sigma_{m}^2)$, and $N\sim \mathcal{N}(\mathbf{0},\sigma_{b}^2\mathbf{I}_d)$ represents the AWGN. During inference phase, this wireless link is mapped to an FC layer operation, thus, ${M'}^{(l_0)}=H$ and ${B'}^{(l_0)}=N$.

A core component of the learning algorithm is the controllable artificial channel during training phase. In this simulation, we evaluate two distinct configurations for the artificial layer $P_{art}^{(l_0)}$. First, in the perfect channel setting, $P_{art}^{(l_0)}$ is an identity mapping $\delta_{(\mathbf{I}_d,\mathbf{0})}$. Second, in the fading channel setting, $M^{(l_0)}$ and $B^{(l_0)}$ are modeled as Gaussian distributions with parameters $(\tilde{\mu}_{m},\tilde{\sigma}_{m})$ and $(0,\tilde{\sigma}_{b})$, respectively. Given that $P_{art}^{(l_0)}$ and $P_{ch}^{(l_0)}$ are continuous and isotropic, their KL divergence can be calculated in a tractable closed-form, which evaluates to zero when their channel statistics are matched perfectly.

We evaluate four training methods to demonstrate the impact of channel awareness. The baselines are standard ERM algorithms, which minimize empirical cross-entropy loss, trained under either a perfect artificial channel (ERM Perfect) or a matched artificial channel where $P_{art}^{(l_0)}=P_{ch}^{(l_0)}$ (ERM Matched). These are compared against our proposed algorithm, similarly trained under perfect (Proposed Perfect, optimizing \eqref{equ:objective_perfect}) and matched (Proposed Matched, optimizing \eqref{equ:objective}) settings, using parameters $\epsilon = 0.025$ and $\eta=0.06$. 
To evaluate the algorithms under varying channel conditions, we establish four distinct inference channel scenarios represented by the tuple $(\mu_{m}, \sigma_{m}, \sigma_{b})$: Scenario 1 (Ideal) at $(1.0, 0.01, 0.01)$; Scenario 2 (Low SNR) at $(1.0, 0.01, 1.0)$; Scenario 3 (Severe Fading) at $(0.5, 1.0, 0.01)$; and Scenario 4 (Severe Fading \& Low SNR) at $(0.5, 1.0, 1.0)$. 
Edge performance is assessed via test accuracy (the 0-1 loss), and the expected population risk using cross-entropy loss. Table \ref{tab:performance} compares these metrics against the expected empirical risk observed during training and the derived upper bound on the wireless generalization error $\Delta$.

\begin{table}[htbp]
\centering
\caption{Inference Performance and Expected Population Risk Bounds}
\label{tab:performance}
\resizebox{\columnwidth}{!}{%
\begin{tabular}{@{}llcccc@{}}
\toprule
\textbf{Channel} & \textbf{Training Method} & \textbf{Test Acc.} & \textbf{Pop. Risk} & \textbf{Emp. Risk} & \textbf{Bound} \\ \midrule
\multirow{4}{*}{\begin{tabular}[c]{@{}l@{}}\textbf{S1} \\ (Ideal)\end{tabular}}
 & ERM (Perfect)                & 0.9335 & 0.1559 & 0.1990 & 1.6376 \\
 & ERM (Matched)                & 0.9285 & 0.1572 & 0.2079 & 3.7318 \\
 & Proposed (Perfect)           & 0.8675 & 0.4517 & 1.9730 & 0.9494 \\
 & Proposed (Matched)            & \textbf{0.9350} & \textbf{0.1598} & 0.2045 & 3.3124 \\ \midrule
\multirow{4}{*}{\begin{tabular}[c]{@{}l@{}}\textbf{S2} \\ (Low SNR)\end{tabular}} 
 & ERM (Perfect)                & 0.6925 & 2.1886 & 0.1990 & 2.2111 \\
 & ERM (Matched)                & 0.8995 & 0.2370 & 0.2754 & 3.9174 \\
 & Proposed (Perfect)           & 0.8550 & 0.5059 & 1.8064 & 7.9356 \\
 & Proposed (Matched)            & \textbf{0.9135} & \textbf{0.2133} & 0.3073 & 3.5480 \\ \midrule
\multirow{4}{*}{\begin{tabular}[c]{@{}l@{}}\textbf{S3} \\ (Severe Fading)\end{tabular}}
 & ERM (Perfect)                & 0.7595 & 0.7113 & 0.1990 & 7.1670 \\
 & ERM (Matched)                & 0.8850 & 0.2781 & 0.3288 & 3.7512 \\
 & Proposed (Perfect)           & 0.8760 & 0.3904 & 1.2861 & 38.1439 \\
 & Proposed (Matched)            & \textbf{0.9005} & \textbf{0.2279} & 0.3199 & 3.4422 \\ \midrule
\multirow{4}{*}{\begin{tabular}[c]{@{}l@{}}\textbf{S4} \\ (Severe Fading \\\& Low SNR)\end{tabular}}
 & ERM (Perfect)                & 0.5825 & 3.2482 & 0.1990 & 7.2014 \\
 & ERM (Matched)                & 0.8400 & 0.3574 & 0.4302 & 3.9087 \\
 & Proposed (Perfect)           & 0.8615 & 0.4239 & 1.2847 & 38.4837 \\
 & Proposed (Matched)            & \textbf{0.8880} & \textbf{0.2914} & 0.4235 & 3.5849 \\ \bottomrule
\end{tabular}%
}
\end{table}
As demonstrated in the table, ignoring channel effects (ERM Perfect) leaves the model vulnerable during wireless edge inference. Indeed, as the channel condition worsens, the test accuracy drops significantly. Incorporating inference channel statistics into ERM training (ERM Matched) yields smaller generalization gaps against unseen channels, although the model still slightly overfits to training channel realizations observed during training. In contrast, our proposed channel-aware algorithm with $P_{art}^{(l_0)}=P_{ch}^{(l_0)}$ directly optimizes for generalization, consistently achieving the highest test accuracy across all scenarios. Finally, though relatively loose, the derived bounds consistently establish a valid upper bound on the expected population risk, validating our PAC-Bayesian framework for edge inference.

\section{Conclusion}
In this paper, we addressed the performance degradation in edge inference systems caused by the inherent stochasticity of wireless channels. By introducing an augmented NN model that integrates channel statistics directly into the weight space, we formalized this performance gap as a wireless generalization error. Utilizing the PAC-Bayesian framework, we derived high-probability bounds that explicitly quantify the penalty induced by channel randomness, thereby providing theoretical guarantees for edge inference. Based on these theoretical bounds, we proposed a channel-aware training algorithm that minimizes a tractable surrogate objective. Simulation results show that the proposed algorithm effectively enhances model robustness and maintains high inference accuracy across different channel conditions compared to the standard ERM.



\begin{thebibliography}{9}
\bibitem{edge-AI}
G.~Zhu, D.~Liu, Y.~Du, C.~You, J.~Zhang, and K.~Huang, ``Toward an intelligent edge: Wireless communication meets machine learning,'' \emph{IEEE Commun. Mag.}, vol.~58, no.~1, pp.~19--25, 2020.

\bibitem{wireless-distributed-learning-survey}
M.~Chen, D.~Gündüz, K.~Huang, W.~Saad, M.~Bennis, A.~V. Feljan, and H.~V. Poor, ``Distributed learning in wireless networks: Recent progress and future challenges,'' \emph{IEEE J. Sel. Areas Commun.}, vol.~39, no.~12, pp.~3579--3605, 2021.

\bibitem{edge-inference}
J.~Shao and J.~Zhang, ``Communication-computation trade-off in resource-constrained edge inference,'' \emph{IEEE Commun. Mag.}, vol.~58, no.~12, pp.~20--26, 2020.




\bibitem{JSCC}
H.~Xie, Z.~Qin, G.~Y. Li, and B.-H. Juang, ``Deep learning enabled semantic communication systems,'' \emph{IEEE Trans. Signal Process.}, vol.~69, pp. 2663--2675, 2021.

\bibitem{JSCC-DeepJSCC}
E.~Bourtsoulatze, D.~B. Kurka, and D.~Gündüz, ``Deep joint source-channel coding for wireless image transmission,'' in \emph{Proc. IEEE Int. Conf. Acoust., Speech, Signal Process. (ICASSP)}, 2019, pp. 4774--4778.

\bibitem{edge-IB}
J.~Shao, Y.~Mao, and J.~Zhang, ``Learning task-oriented communication for edge inference: An information bottleneck approach,'' \emph{IEEE J. Sel. Areas Commun.}, vol.~40, no.~1, pp.~197--211, 2022.


\bibitem{JSCC-multilevel}
X.~Gao, H.~Yin, Y.~Sun, D.~Wei, X.~Xu, H.~Chen, W.~Wu, and S.~Cui, ``Multilevel feature transmission in dynamic channels: A semantic knowledge base and deep-reinforcement-learning-enabled approach,'' \emph{IEEE Internet Things J.}, vol.~12, no.~8, pp. 10\,150--10\,162, 2025.




\bibitem{JSCC-VQ}
M.~Lokumarambage, T.~Sivalingam, H.~Rezaei, P.~Rajatheva, and A.~Fernando, ``Edge computing based vector quantized semantic communication system for wireless image transmission,'' \emph{IEEE Trans. Consum. Electron.}, pp. 1--1, 2026.

\bibitem{blind-JSCC}
H.~Yuan, W.~Xu, Y.~Wang, and X.~Wang, ``Channel-blind joint source–channel coding for wireless image transmission,'' \emph{Sensors}, vol.~24, no.~12, 2024.



\bibitem{JSCC-attention}
J.~Xu, B.~Ai, W.~Chen, A.~Yang, P.~Sun, and M.~Rodrigues, ``Wireless image transmission using deep source channel coding with attention modules,'' \emph{IEEE Trans. Circuits Syst. Video Technol.}, vol.~32, no.~4, pp. 2315--2328, 2022.

\bibitem{JSCC-multitask}
X.~Chen, S.~Gan, C.~Feng, X.~Wang, and T.~Q. Quek, ``Take what you need: Flexible multi-task semantic communications with channel adaptation,'' \emph{arXiv preprint arXiv:2502.08221}, 2025.






\bibitem{compression}
Y.~Bu, W.~Gao, S.~Zou, and V.~V. Veeravalli, ``Population risk improvement with model compression: An information-theoretic approach,'' \emph{Entropy}, vol.~23, no.~10, 2021.

\bibitem{variable-compression}
M.~Sefidgaran and A.~Zaidi, ``Data-dependent generalization bounds via variable-size compressibility,'' in \emph{Proc. IEEE Int. Symp. Inf. Theory (ISIT)}, 2024, pp. 2682--2687.

\bibitem{pac-bayes-catoni}
O.~Catoni, ``A pac-bayesian approach to adaptive classification,'' \emph{preprint}, vol. 840, no.~2, pp.~6, 2003.

\bibitem{pac-bayes-mcallester}
D.~A. McAllester, ``Pac-bayesian model averaging,'' in \emph{Proc. Annu. Conf. Comput. Learn. Theory (COLT)}, 1999, pp.~164--170. 

\bibitem{proof}
Y.~He, G.~Yu, and J.~Zhu, ``A pac-bayesian analysis of channel-induced degradation in edge inference,'' \emph{arXiv preprint}, 2026.

\bibitem{genearlization-bound-book}
F.~Hellstr{\"o}m, G.~Durisi, B.~Guedj, and M.~Raginsky, ``Generalization bounds: Perspectives from information theory and pac-bayes,'' \emph{Foundations and Trends in Machine Learning}, vol.~18, no.~1, pp. 1--223, 2025.

\bibitem{BBB}
C.~Blundell, J.~Cornebise, K.~Kavukcuoglu, and D.~Wierstra, ``Weight uncertainty in neural networks,'' in \emph{Proc. Int. Conf. Mach. Learn. (ICML)}, 2015, vol. 37, pp.~1613--1622.

\bibitem{PBB}
M.~P\'{e}rez-Ortiz, O.~Rivasplata, J.~Shawe-Taylor, and C.~Szepesv\'{a}ri, ``Tighter risk certificates for neural networks,'' \emph{J. Mach. Learn. Res.}, vol.~22, no.~1, Jan. 2021.

\bibitem{two-moons}
T.~B\"{u}hler and M.~Hein, ``Spectral clustering based on the graph p-laplacian,'' in \emph{Proc. Int. Conf. Mach. Learn. (ICML)}, 2009, p. 81–88.


\end{thebibliography}





\appendices

\section{Donsker-Varadhan variational representation}
\begin{lemma}
\label{thm:Donsker_Varadhan}
For any two probability measures $\rho$ and $\pi$ defined on a measurable space $(\mathcal{X},\Sigma)$ where $\Sigma$ is a $\sigma$-algebra on the set $\mathcal{X}$. Let $f:\mathcal{X}\to \R$ be a measurable function such that $\E_{X\sim\pi}\left[e^{f(X)}\right]<\infty$, then
\begin{equation}
  \label{equ:Donsker_Varadhan}
  D(\rho\|\pi) = \sup_f\left\{\E_{X\sim\rho}\left[f(X)\right] - \log \E_{X\sim\pi}\left[e^{f(X)}\right]\right\}.
\end{equation}
\end{lemma}

\section{Proof of Theorem \ref{thm:simple_channel_bound}}
\label{apx:proof_theorem_1}
\begin{proof}
The wireless generalization error can be decomposed into two components
\begin{equation}
    \label{equ:Delta_decompose}
    \begin{split}
        \Delta & = \E_{P_{W'|S}}\E_{P_Z}\left[l\left(W',Z\right)\right] - \E_{P_{W|S}}\left[\frac{1}{n}\sum_{i=1}^nl(W,Z_i)\right] \\
        & = \underbrace{\E_{P_{W'|S}P_Z}\left[l\left(W',Z\right)\right] - \E_{P_{W|S}P_Z}\left[l\left(W,Z\right)\right]}_A \\
        & + \underbrace{\E_{P_{W|S}}\E_{P_Z}\left[l\left(W,Z\right)\right] - \E_{P_{W|S}}\left[\frac{1}{n}\sum_{i=1}^nl(W,Z_i)\right]}_B.
    \end{split}
\end{equation}
Term $B$ can be upper bounded by applying Donsker-Varadhan as in conventional PAC-Bayesian framework \cite{genearlization-bound-book}: If $l(w,z)$ is $\sigma$-sub-Gaussian over $P_Z$ for all $w\in\mathcal{W}$, then for all data-independent prior $Q_W$ over $\mathcal{W}$ and $k>0$,
\begin{equation}
    \Pr_S\left\{B \leq \sqrt{\frac{2\sigma^2}{n-1}\left(D(P_{W|S}\|Q_W\right)+\log\frac{\sqrt{n}}{\epsilon}}\right\} > 1-\epsilon.
\end{equation}

Term $A$ is equivalent to
\begin{equation}
    A = \E_{P_Z}\left[\E_{P_{W'|S}}\left[l\left(W',Z\right)\right] - \E_{P_{W|S}}\left[l\left(W,Z\right)\right]\right].
\end{equation}
The difference $\E_{P_{W'|S}}[l(W',Z)] - \E_{P_{W|S}}[l(W,Z)]$ can be upper bounded as in \eqref{equ:bound_A} for some $\lambda>0$, where (a) is utilizing Donsker-Varadhan lemma with $f=-\lambda l(W',Z)$, $\rho=P_{W|S}$, and $\pi=P_{W'|S}$ and (b) is from the assumption that the loss function $l\left(W,Z\right)$ is $\sigma_0$-sub-Gaussian under $P_{W'|S}$ for all $S$. 
\begin{figure*}[tb]
\begin{equation}
    \label{equ:bound_A}
    \begin{split}
        & \E_{P_{W'|S}}\left[l\left(W',Z\right)\right] - \E_{P_{W|S}}\left[l\left(W,Z\right)\right] \\
        & = \frac{1}{\lambda}\left(\E_{P_{W'|S}}\left[\lambda l\left(W',Z\right)\right] + \E_{P_{W|S}}\left[-\lambda l\left(W,Z\right)\right]\right) \\
        & \overset{(a)}{\leq} \frac{1}{\lambda}\left(\E_{P_{W'|S}}\left[\lambda l\left(W',Z\right)\right] + D(P_{W|S}\|P_{W'|S}) + \log \E_{P_{W'|S}}\left[e^{-\lambda l\left(W',Z\right)}\right]\right) \\
        & = \frac{1}{\lambda}\left(\E_{P_{W'|S}}\left[\lambda l\left(W',Z\right)\right] + D(P_{W|S}\|P_{W'|S}) + \log \E_{P_{W'|S}}\left[e^{-\lambda \left(l\left(W',Z\right)-\E_{P_{W'|S}}\left[l\left(W',Z\right)\right]\right) - \lambda \E_{P_{W'|S}}\left[l\left(W',Z\right)\right]}\right]\right) \\
        & = \frac{1}{\lambda}\left(\E_{P_{W'|S}}\left[\lambda l\left(W',Z\right)\right] + D(P_{W|S}\|P_{W'|S}) + \log \E_{P_{W'|S}}\left[e^{-\lambda \left(l\left(W',Z\right)-\E_{P_{W'|S}}\left[l\left(W',Z\right)\right]\right)}e^{-\lambda \E_{P_{W'|S}}\left[l\left(W',Z\right)\right]}\right]\right) \\
        & = \frac{1}{\lambda}\left(\E_{P_{W'|S}}\left[\lambda l\left(W',Z\right)\right] + D(P_{W|S}\|P_{W'|S}) + \log\left(\E_{P_{W'|S}}\left[e^{-\lambda \left(l\left(W',Z\right)-\E_{P_{W'|S}}\left[l\left(W',Z\right)\right]\right)}\right]e^{-\lambda \E_{P_{W'|S}}\left[l\left(W',Z\right)\right]}\right)\right) \\
        & = \frac{1}{\lambda}\left(\E_{P_{W'|S}}\left[\lambda l\left(W',Z\right)\right] + D(P_{W|S}\|P_{W'|S}) + \Lambda_{l\left(W',Z\right)}^{P_{W'|S}}(-\lambda) + \log\left(e^{-\lambda \E_{P_{W'|S}}\left[l\left(W',Z\right)\right]}\right)\right) \\
        & = \frac{1}{\lambda}\left(\E_{P_{W'|S}}\left[\lambda l\left(W',Z\right)\right] + D(P_{W|S}\|P_{W'|S}) + \Lambda_{l(W,Z)}^{P_{W'|S}}(-\lambda) - \lambda \E_{P_{W'|S}}\left[l\left(W',Z\right)\right]\right) \\
        & = \frac{1}{\lambda}\left(D(P_{W|S}\|P_{W'|S}) + \Lambda_{l(W,Z)}^{P_{W'|S}}(-\lambda)\right) \\
        & \overset{(b)}{\leq} \frac{1}{\lambda}\left(D(P_{W|S}\|P_{W'|S}) + \frac{\lambda^2\sigma_0^2}{2} \right) \\
        & = \frac{D(P_{W|S}\|P_{W'|S})}{\lambda} + \frac{\lambda\sigma_0^2}{2}
    \end{split}
\end{equation}
\end{figure*}
Since this inequality holds for all dataset $S$, we can minimize the RHS over $\lambda$ and obtain
\begin{equation}
    A \leq \sqrt{2\sigma_0^2D(P_{W|S}\|P_{W'|S})}.
\end{equation}

Combining these results yields that w.p. at least $1-\epsilon$, we have
\begin{equation}
    \label{equ:no_lip_old_channel}
    \begin{split}
        \Delta &\leq \sqrt{2\sigma_0^2D(P_{W|S}\|P_{W'|S})}  \\
        &+ \sqrt{\frac{2\sigma^2}{n-1}\left(D(P_{W|S}\|Q_W\right)+\log\frac{\sqrt{n}}{\epsilon}}.
    \end{split}
\end{equation}

Utilizing the definition for $P_{W|S}$ in \eqref{equ:learning_algorithm} and $P_{W'|S}$ in \eqref{equ:P_W'|S}, and let the data-independent prior $Q_W$ be defined as
\begin{equation}
    Q_W(w) = Q_{\tilde{W}}\left(\tilde{w}\right)P_{art}^{(l_0)}\left(w^{(l_0)}\right),
\end{equation}
where $Q_{\tilde{W}}$ is any data-independent prior, the KL divergence $D(P_{W|S}\|P_{W'|S})$ is equivalent to
\begin{equation}
    \begin{split}
        & D(P_{W|S}\|P_{W'|S}) \\
        & = \E\left[\log\frac{P_{W|S}(w|s)}{P_{W'|S}(w|s)}\right] \\
        & = \E\left[\log\frac{P_{\tilde{W}|W^{(l_0)}S}\left(\tilde{w}|w^{(l_0)},s\right)P_{art}^{(l_0)}\left(w^{(l_0)}\right)}{P_{\tilde{W}|S}\left(\tilde{w}|s\right)P_{ch}^{(l_0)}\left({w}^{(l_0)}\right)}\right] \\
        & = \E\left[\log\frac{P_{\tilde{W}|W^{(l_0)}S}\left(\tilde{w}|w^{(l_0)},s\right)}{P_{\tilde{W}|S}\left(\tilde{w}|s\right)}\right] + \E\left[\log\frac{P_{art}^{(l_0)}\left(w^{(l_0)}\right)}{P_{ch}^{(l_0)}\left({w}^{(l_0)}\right)}\right] \\
        & = I_S\left(\tilde{W};W^{(l_0)}\right) + D\left(P_{art}^{(l_0)}\big\|P_{ch}^{(l_0)}\right),
    \end{split}
\end{equation}
where mutual information is defined as $I_S(\tilde{W};W^{(l_0)})=\E_{P_{art}^{(l_0)}}[D(P_{\tilde{W}|W^{(l_0)}S}\|P_{\tilde{W}|S})]$. Meanwhile, the KL divergence $D(P_{W|S}\|Q_W)$ can be simplified to
\begin{equation}
    \begin{split}
        & D(P_{W|S}\|Q_W) \\
        & = \E\left[\log\frac{P_{W|S}(w|s)}{Q_W(w)}\right] \\
        & = \E\left[\log\frac{P_{\tilde{W}|W^{(l_0)}S}\left(\tilde{w}|w^{(l_0)},s\right)P_{art}^{(l_0)}\left(w^{(l_0)}\right)}{Q_{\tilde{W}}\left(\tilde{w}\right)P_{art}^{(l_0)}\left(w^{(l_0)}\right)}\right] \\
        & = \E\left[\log\left(\frac{P_{\tilde{W}|W^{(l_0)}S}\left(\tilde{w}|w^{(l_0)},s\right)}{Q_{\tilde{W}}\left(\tilde{w}\right)}\right)\right] \\
        & = \E_{P_{art}^{(l_0)}}\left[D\left(P_{\tilde{W}|W^{(l_0)}S}\|Q_{\tilde{W}}\right)\right].
    \end{split}
\end{equation}
Summarizing the results yields: for all data-independent prior $Q_{\tilde{W}}$, w.p. at least $1-\epsilon$ over $S$, we have
\begin{equation}
    \begin{split}
        & \Delta\leq \sqrt{2\sigma_0^2\left(I_S\left(\tilde{W};W^{(l_0)}\right) + D\left(P_{art}^{(l_0)}\big\|P_{ch}^{(l_0)}\right)\right)} \\
        & + \sqrt{\frac{2\sigma^2}{n-1}\left(\E_{P_{art}^{(l_0)}}\left[D\left(P_{\tilde{W}|W^{(l_0)}S}\|Q_{\tilde{W}}\right)\right]+\log\frac{\sqrt{n}}{\epsilon}\right)}.
    \end{split}
\end{equation}
\end{proof}

\section{Proof of Theorem \ref{thm:perfect_training_simple_channel_bound}}
\label{apx:proof_theorem_2}
The wireless generalization error $\Delta$ can be decomposed into $A$ and $B$, as in \eqref{equ:Delta_decompose}.
Term $A$ can be upper bounded under the Lipschitz assumption:
\begin{equation}
    \begin{split}
        A & = \E_{P_{W'|S}}\E_{P_Z}\left[l\left(W',Z\right)\right] - \E_{P_{W|S}}\E_{P_Z}\left[l\left(W,Z\right)\right] \\
        & = \E_{P_{W'W|S}\E_{P_Z}}\left[l\left(W',Z\right) - l(W,Z)\right] \\
        & \overset{(a)}{\leq} \E_{P_{W'W|S}\E_{P_Z}}\left[Kd_{\mathcal{W}}\left(W',W\right)\right] \\
        & = \E_{P_{W'W|S}}\left[Kd_{\mathcal{W}}\left(W',W\right)\right] \\
        & \overset{(b)}{=} \E\left[K\left(d_{\mathcal{\tilde{W}}}\left(\tilde{W}',\tilde{W}\right) + d^{(l_0)}\left({W'}^{(l_0)},W^{(l_0)}\right)\right)\right] \\
        & \overset{(c)}{=} \E\left[K d^{(l_0)}\left({W'}^{(l_0)},W^{(l_0)}\right)\right] \\
        & \overset{(d)}{=} \E_{{w'}^{(l_0)}\sim P_{ch}^{(l_0)}}\left[K\left\|{w'}^{(l_0)}-(\mathbf{I},\mathbf{0})\right\|_F\right] ,
    \end{split}
\end{equation}
where (a) is from the loss function $l(w,z)$ is $K$-Lipschitz continuous on $\mathcal{W}$ for all $z\in\mathcal{Z}$, (b) is from the layer-wise distance metric $d_{\mathcal{W}}(W',W) = \sum_{l=1}^{L+1}d^{(l)}({W'}^{(l)},{W}^{(l)})$, (c) and (d) are from the definition of joint distribution $P_{W'W|S}=P_{W'|W}P_{W|S}$.

Term $B$ can be upper bounded by applying Donsker-Varadhan as in conventional PAC-Bayesian framework \cite{genearlization-bound-book}: If $l(w,z)$ is $\sigma$-sub-Gaussian over $P_Z$ for all $w\in\mathcal{W}$, then for all data-independent prior $Q_W$ over $\mathcal{W}$ and $k>0$,
\begin{equation}
    \Pr_S\left\{B \leq \sqrt{\frac{2\sigma^2}{n-1}\left(D(P_{W|S}\|Q_W\right)+\log\frac{\sqrt{n}}{\epsilon}}\right\} > 1-\epsilon.
\end{equation}
Summarizing the results completes the proof.

\section{Proof of Corollary \ref{thm:perfect_training_simple_channel_bound_relaxed}}
\label{apx:proof_corollary_1}
Applying Donsker-Varadhan on the channel penalty term with $f=K\|{W'}^{(l_0)}-(\mathbf{I},\mathbf{0})\|_F$, $\rho=P_{ch}^{(l_0)}$, and $\pi=Q^{(l_0)}$, the result follows immediately:
\begin{equation}
    \begin{split}
        & \E_{P_{ch}^{(l_0)}}\left[K\left\|{W'}^{(l_0)}-(\mathbf{I},\mathbf{0})\right\|_F\right] \\
        & \leq \log \E_{Q^{(l_0)}}\left[e^{K\left\|{W'}^{(l_0)}-(\mathbf{I},\mathbf{0})\right\|_F}\right] + D\left(P_{ch}^{(l_0)}\|Q^{(l_0)}\right).
    \end{split}
\end{equation}

\end{document}